 \UseRawInputEncoding
 
 \documentclass[conference]{IEEEtran}
\IEEEoverridecommandlockouts
\usepackage{cite}
\usepackage{amsmath,amssymb,amsfonts}
\usepackage{algorithmic}
\usepackage{graphicx}
\usepackage{booktabs}
\usepackage{textcomp}
\usepackage{xcolor}
\usepackage{siunitx}
\usepackage{soul}
\def\BibTeX{{\rm B\kern-.05em{\sc i\kern-.025em b}\kern-.08em
    T\kern-.1667em\lower.7ex\hbox{E}\kern-.125emX}}
\usepackage{graphicx}

\usepackage[hyphens]{url}

\begin{document}

\title{Ultrawideband USRP-Based Channel Sounding Utilizing the RFNoC Framework\\
}

\author{
\IEEEauthorblockN{Michiel Sandra, Christian Nelson, Anders J Johansson}
\IEEEauthorblockA{Electrical and Information Technology \\
\textit{Lund University}\\
Lund, Sweden \\
\{michiel.sandra,christian.nelson,anders\_j.johansson\}@eit.lth.se}



}

\maketitle

\begin{abstract}
This paper shows how an ultrawideband channel sounder can be built with the latest National Instruments (NI) Universal Software Radio Peripheral (USRP) equipment, featuring onboard FPGA processing and utilizing open-source tools. Compared to other USRP-based channel sounders, our design is versatile, eases the requirements on the data link between the USRPs and the host, limits the measurement data size, and increases the signal-to-interference-plus-noise ratio due to averaging. Furthermore, this paper serves as a proof-of-concept for other researchers seeking to build a channel sounder based on the USRP equipment.
\end{abstract}
\section{Introduction}


New parts of the spectrum are available with the new fifth and sixth generation wireless standards. The systems go higher in frequency and are also becoming more wide-band, requiring us to deepen our knowledge about wireless channels. Measurement of wireless channels, also known as channel sounding, has played a vital role in developing realistic channel models.
In the last decade, channel sounding based on software-defined radios (SDR), such as the Universal Software Radio Peripheral (USRP) platform from National Instruments (NI), has gained much popularity due to flexibility, portability, and cost-efficiency \cite{Merwaday2014, Zelenbaba2020, Stanko2021, Zhang2020,8576578,9454158,7928479}.
With the third generation NI \mbox{USRPs}, the bandwidth has been limited to 160 MHz. 
The release of the new X410 USRP has now opened up for bandwidths up to 400 MHz, which can provide a more detailed insight into the wireless propagation channels.
However, such a large bandwidth imposes challenges in handling the massive stream of I/Q samples. 
Whereas most USRP-based channel sounder designs stream all sample data to the host computer at the receiver side, 
we have developed a custom signal processing block on the FPGA onboard the
X410 to limit the data stream. The authors of \cite{similar_paper} also developed a USRP-based channel sounder containing custom signal processing blocks; however, our design is more compact and and can handle a higher bandwidth.
Our custom signal processing block was developed within the open-source framework RF Network on Chip (RFNoC) by Ettus Research~\cite{rfnoc}. 
This framework is responsible for the streaming and processing of RF data on the FPGA, and offers the possibility to create custom applications based on so-called RFNoC blocks. There are already various types of blocks available; nevertheless, one can create one's own block tailored to a specific purpose. 

\begin{figure}
    \centering
    \includegraphics[width=\columnwidth]{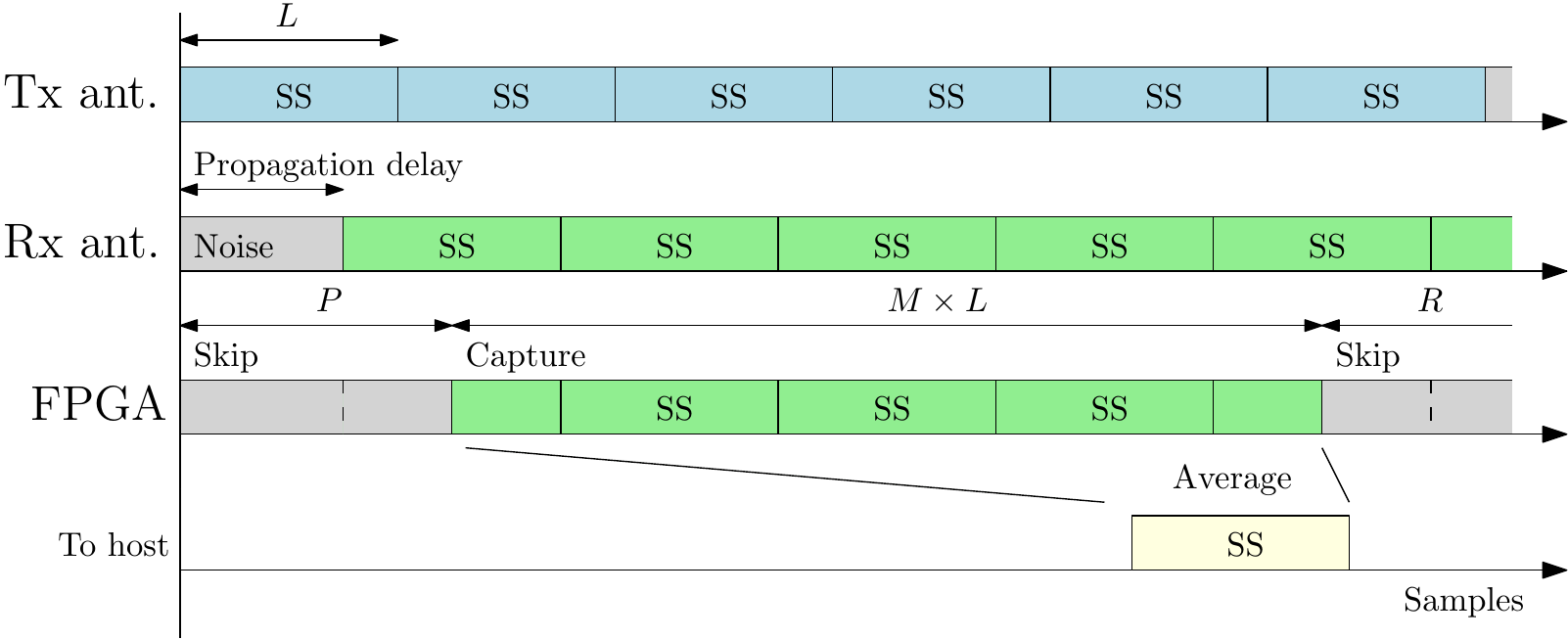}
    \caption{RFNoC processing block functionality for one channel snapshot}
    \label{fig:sounder}
\end{figure}

\section{Implementation}
The transmitter and the receiver of the channel sounder each consist of a host computer and a NI USRP X410, which is a four-channel full-duplex SDR based on the Xilinx UltraScale+ RFSoC. The center frequency can be tuned from 1\,MHz up to 7.2 GHz, and bandwidths up to 400\,MHz are supported. 
Custom development has been done for the host computer and for the FPGA on the X410.
In the following paragraphs, we first give a functional description of the channel sounder, followed by the details of the development on FPGA and the host computer. In the last subsection, we also discuss how the transmitter and the receiver are synchronized. 

\subsection{Functional Description}
During one channel snapshot, the transmitter sends multiple repetitions of the
same sounding signal (SS) with a length of $L$ samples, as illustrated in Fig.~\ref{fig:sounder}.
Note that to avoid aliasing $L$ should be longer than
\begin{equation}
    L > \frac{\Delta\tau_\mathrm{max}}{T_s}
\end{equation}
where $\Delta\tau_\mathrm{max}$ is the maximum delay difference between the first and last received multi-path component, and $T_s$ is the sample period.
These signals then propagate through the wireless channel and arrive at
the receiver with an unknown propagation delay. Hence, samples have to be discarded according to this propagation delay. Otherwise, a couple of samples will be averaged with noise, which reduces the signal-to-noise ratio. In addition, another $L$ samples have to be discarded so the received sounding signals are the result of a circular convolution between the channel impulse response and the original sounding signal. We define $P$ as the total number of samples that are discarded and it should be higher than
\begin{equation}
    P \ge \frac{\tau_\mathrm{0,max}}{T_s} + L 
\end{equation}
where $\tau_\mathrm{0,max}$ the maximum delay of the first multi-path component.
In practice, $P$ can be set higher than required to account for eventual timing synchronization errors between receiver and transmitter.
After disregarding $P$ samples, $M$ signals are captured and averaged. 
The averaging is done by averaging each $i^\mathrm{th}$ sample of each received sounding signal where $i \in [0,L-1]$, resulting in averaged received sounding signal of length $L$, as depicted in Fig.~\ref{fig:sounder}.
Then, the averaged signal is sent to the host for channel estimation and storage. The rest of the received samples are skipped until the following channel snapshot.

\begin{figure}
    \centering
    \includegraphics[width=\columnwidth]{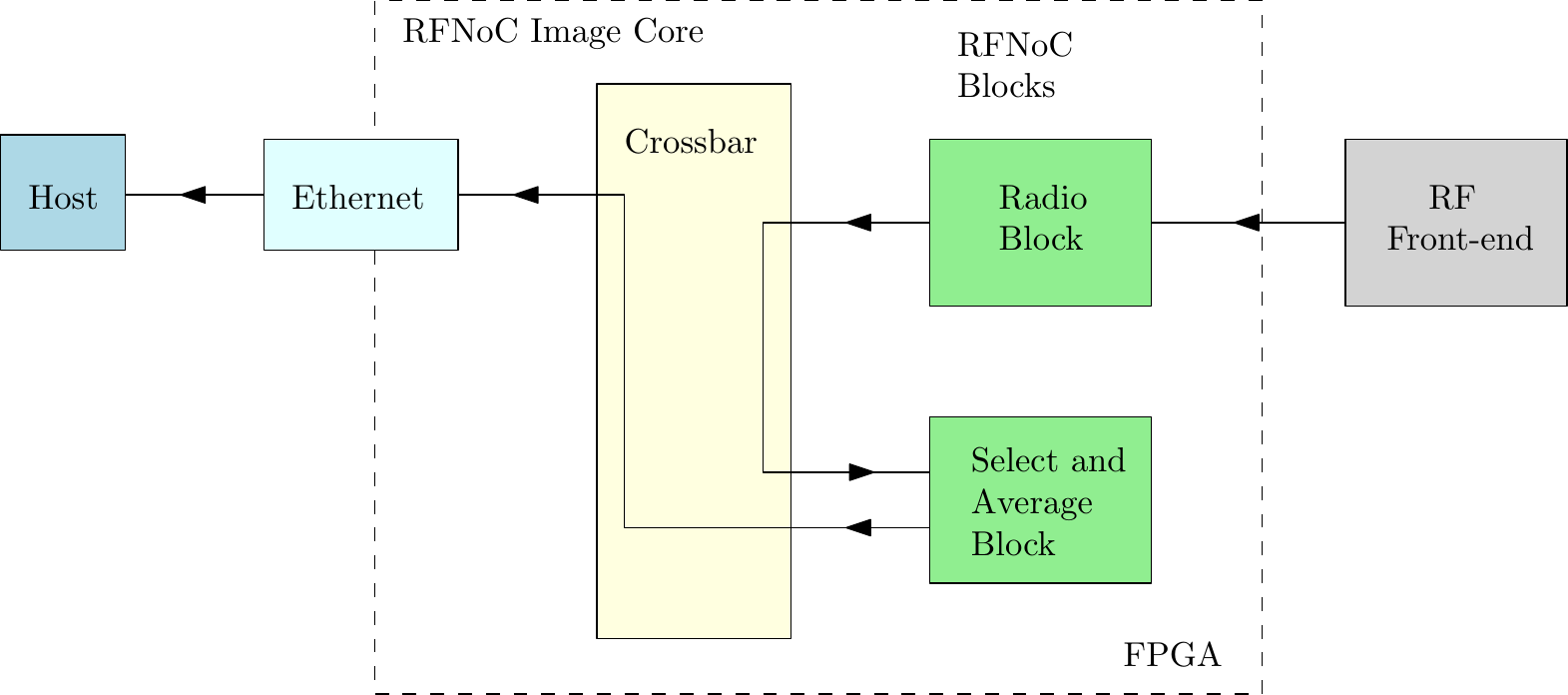}
    \caption{RFNoC architecture at the receiver}
    \label{fig:arch}
\end{figure}

\subsection{Logic circuit on FPGA}

At the \textit{transmitter}, we utilized the X4\_400 FPGA image flavor, part of the UHD Release v4.2.0.0 \cite{uhd_github}, which includes a four-channel replay block. With the replay block, we can download our sounding signal into the DRAM of the X410 and then transmit it at the appropriate time. The sounding signal loaded into the DRAM of the transmitter has the length of the channel measurement interval $T_\mathrm{rep}$, where the first part contains the $\lceil\frac{M L + P}{L}
\rceil$ sounding signals, and the second part is filled with zeros. By using the onboard DRAM to transmit the sounding signal, there is no transfer of sample data between the host computer and the USRP while the USRP is transmitting. Hence, a standard Ethernet cable can be used to connect the USRP and the host computer. 

\begin{figure}
    \centering
    \includegraphics[width=0.8\columnwidth]{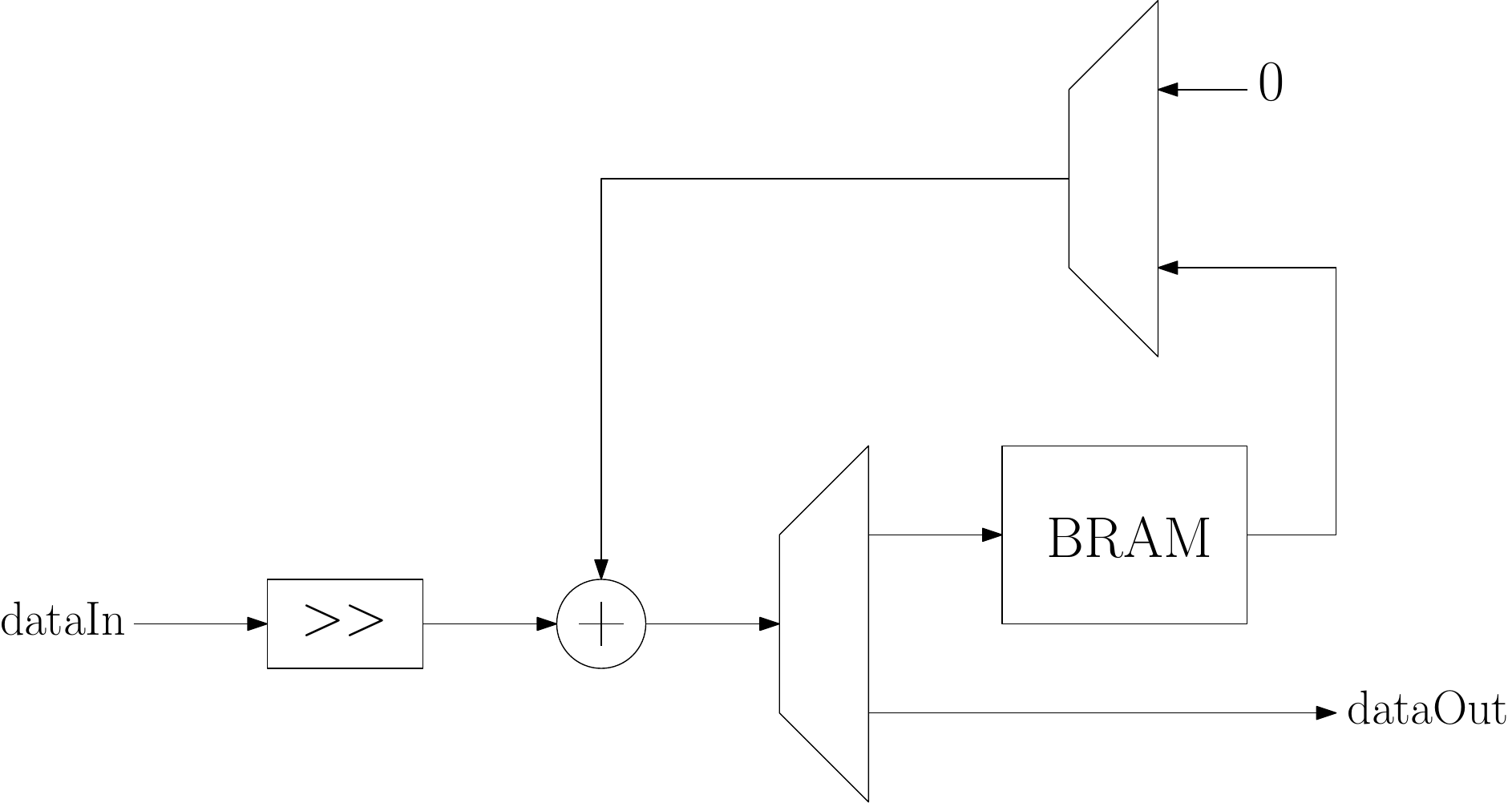}
    \caption{Simplified top-level circuit diagram of the averager in the select and average RFNoC block.}
    \label{fig:circuit}
\end{figure}

At the \textit{receiver}, we took the same flavor as a basis but installed our custom RFNoC block, written in Verilog, replacing the replay RFNoC block. The high-level architecture of the receiver FPGA is depicted in Fig.~\ref{fig:arch}. The custom block selects samples based on a sample counter and averages the received sounding signals described in the previous subsection. 
Since the data type is complex short, the averaging could be efficiently done by bit-wise right shifting each sample $K$ times and then adding up the $i^\mathrm{th}$ sample of each sounding signal. The top-level circuit diagram of the averager is depicted in Fig.~\ref{fig:circuit}. The averager consists of a shifter, block RAM (BRAM), and an adder. The multiplexers are controlled by a state machine which consist of three states: 1) \textit{IN}: The samples are saved in the BRAM. 2) \textit{ADD\_IN}: The $i^\mathrm{th}$ sample comes in and the $i^\mathrm{th}$ sample is fetched out of the memory, and their sum is stored in the memory. 3) \textit{ADD\_OUT}: The $i^\mathrm{th}$ sample comes in and the $i^\mathrm{th}$ sample is fetched out of the memory. The sum of those samples is presented at the output of the averager.
Because of the high sample rate, two samples are being processed per clock cycle, which results in a memory width of 64 bits. The clock driving this RFNoC block is \texttt{clk\_radio2x}, which is equal to of 250\,MHz.

\begin{table}[]
\centering
\caption{Overview of the parameters of the channel sounder \label{tab:param}}
\begin{tabular}{@{}ccl@{}}
\toprule
\textbf{Parameter} & \textbf{Value} & \textbf{Description}\\ \midrule
$L$ & 1024 samples & Sound signal length\\
$P$ & 2048 samples & Skipped samples due to propagation\\
$M$ & 64 signals   & Average quantity \\
$K$ & 6 bits & Bit shift in averager \\
$R$ & 2432416 samples & Skipped samples until next snapshot\\
$f$ & 5.725\,GHz &  Center frequency \\
$P_\mathrm{Tx}$ & 14 dBm & Output power \\
$T_\mathrm{rep}$ & 5 ms & Channel measurement repetition \\ 
$T_s$ & 2 ns & Sample period \\ 
\bottomrule

\end{tabular}
\end{table}

\subsection{Software on host computer}
A custom C++ application has been written which utilizes the UHD library (Release v4.2.0.0)\cite{uhd_github}, containing all the essential software to control the USRPs. An additional library was also developed to set the parameters ($L$, $P$, $K$, and $M$) belonging to the RFNoC block. 

\subsection{Synchronization of transmitter and receiver}
The transmitter and the receiver USRPs are synchronized via a PPS signal for time synchronization, and a 10 MHz reference signal for frequency synchronization. Through the main application on the host computer, the synchronization source can be set to either a global positioning system (GPS) disciplined oscillator or an external clock (e.g. Rubidium clock).
The channel sounder is designed to start at the positive flank of a PPS. In order to remove the requirement for the transmitter and the receiver to start at the same PPS flank, the repetition time $T_\mathrm{rep}$ needs to be chosen in such a way that at each PPS flank a sounding signal starts transmitting, e.g. $T_\mathrm{rep}$~=~5 ms. In this way, the transmitter and receiver can start at different PPS flanks; hence, no communication is needed between the transmitter and the receiver. Nevertheless, the transmitter and receiver can also start on the same PPS flank via the GPS time when the system is synchronized with GPS.

\begin{figure}
    \centering
    \includegraphics[width=0.8\columnwidth]{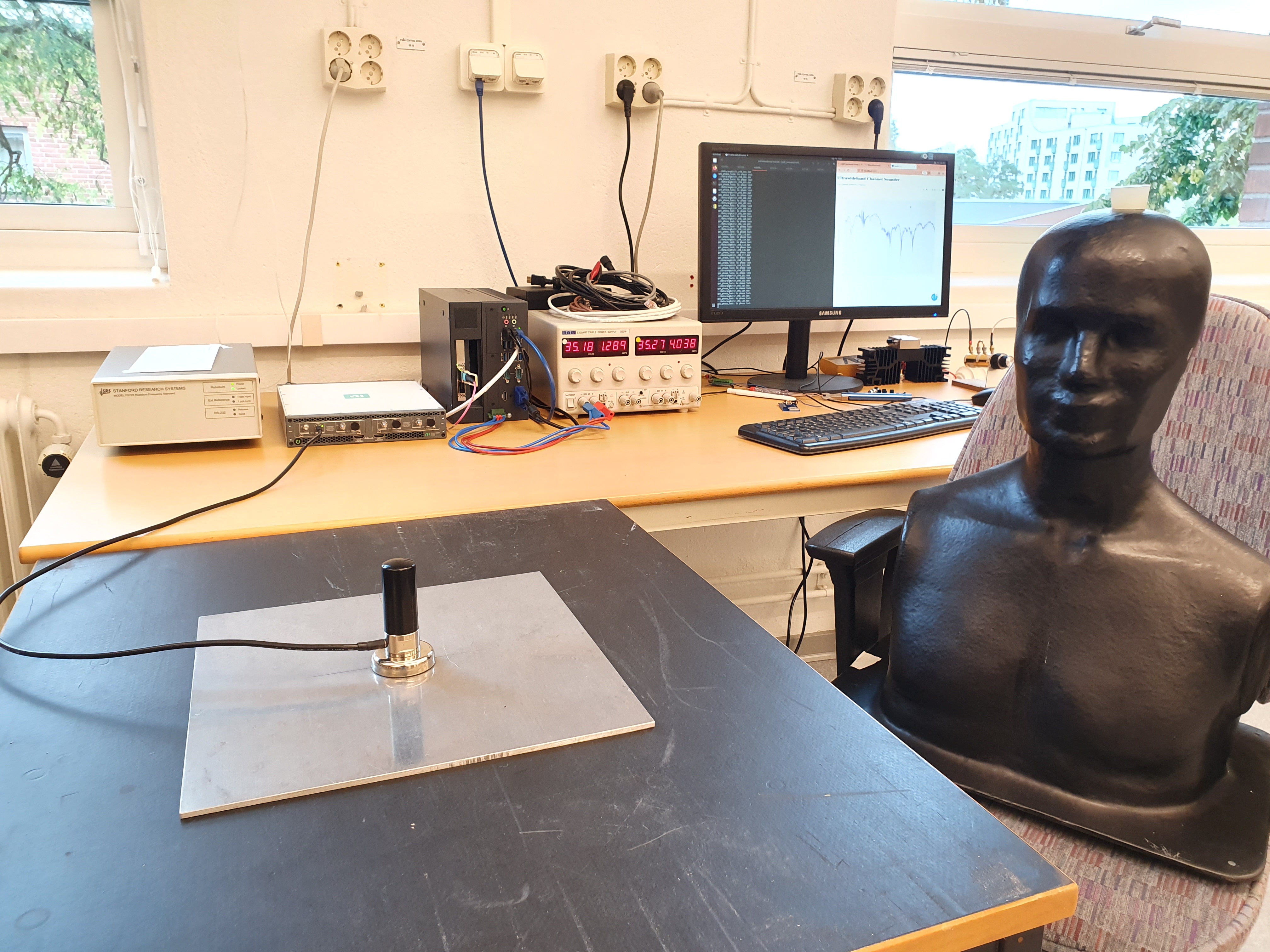}
    \caption{Measurement setup on the receiver side. Equipment on the desk from left to rigth: Rubidium clock, NI USRP X410, PC (Advantech  MIC-770 V2), a power supply for the PC.}
    \label{fig:setup}
\end{figure}

\begin{figure}
    \centering
    \includegraphics[width=\columnwidth]{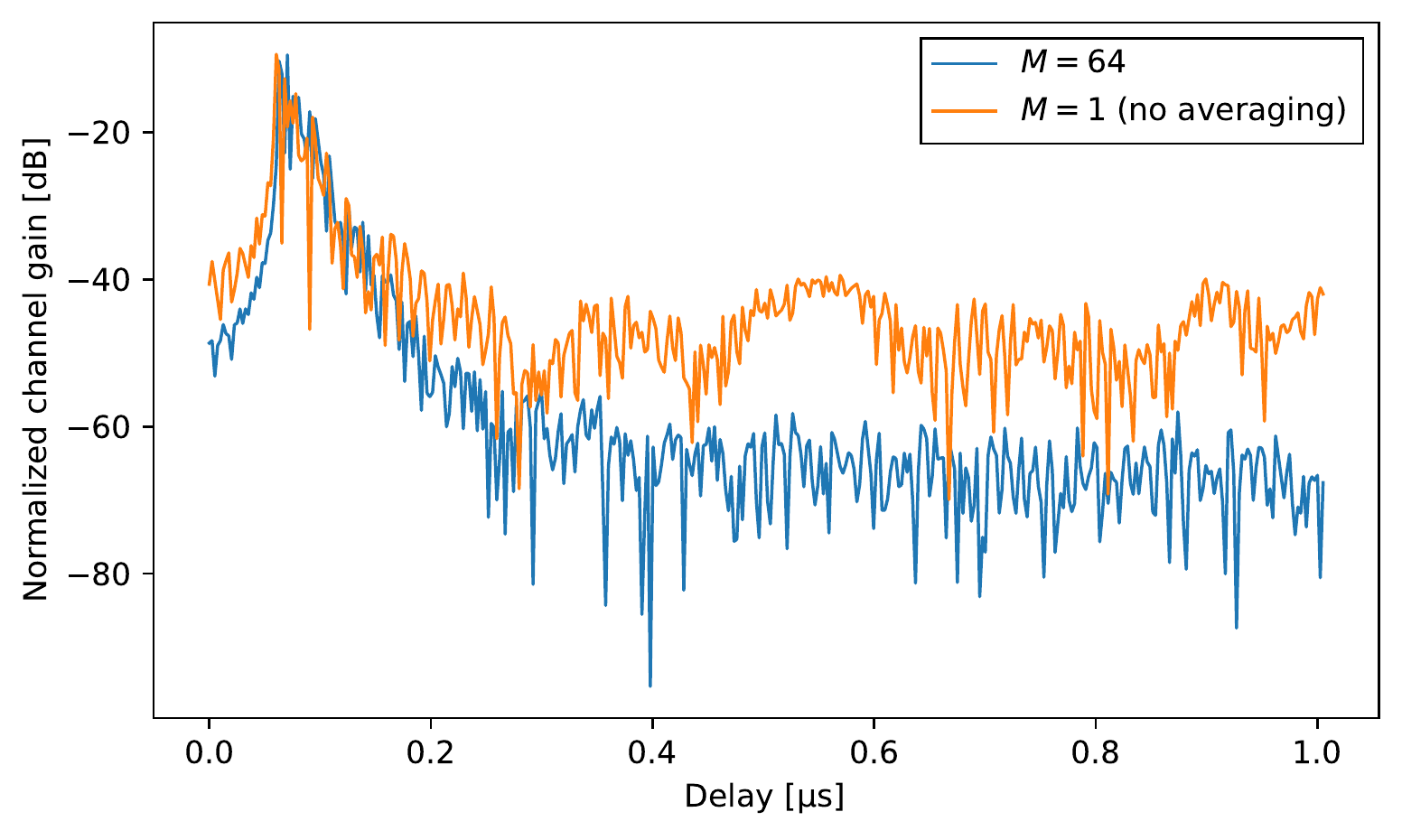}
    \caption{Measured channel impulse responses in a lab environment at 5.725\,GHz with 400\,MHz bandwidth, with and without averaging.}
    \label{fig:cir}
\end{figure}

\section{Evaluation}
Our design has been validated by employing a test-bench that verified sample selection and averaging. In addition, we performed single-input, single-output channel measurements of an indoor lab environment with two omnidirectional antennas at 5.725\,GHz separated by 5 meters. A picture of our setup can be found in Fig.~\ref{fig:setup}. The sounding signal, a Zadoff-Chu (ZC) signal, was sent in an OFDM scheme, covering 813 out of 1024 sub-carriers, so that it covers a bandwidth of about 400\,MHz at a sample rate of 500\,Msps. The parameter configuration of the channel can be found in Table~\ref{tab:param}. In Fig.~\ref{fig:cir}, the measured channel impulse response is depicted for two measurements, one with averaging and one without averaging. Several multi-path components can be distinguished in the figure due to the ultra-wide bandwidth. For $M=1$, one can observe a higher noise level and several lobes beyond \SI{0.2}{\micro\second}. These lobes are caused by in-band interference coming from an unknown device. However, by averaging the received signals, this interference can be suppressed. Also, the effects of the RF chains are taken into account by means of a back-to-back calibration. Because of the onboard processing, the average data rate to the host computer during the measurement was reduced to about 1~MB/s.

\section{Conclusion}
In this paper, we have presented the design of our UWB channel sounder built with the USRP X410 and open-source tools that can be employed in various channel measurement scenarios. Owing to a custom RFNoC block on the FPGA that performs sample selection and averaging, the data rate to the host computer could be reduced significantly. In the future, the design will be extended so it is capable of measuring multiple-input multiple-output channels.

\bibliographystyle{IEEEtran}
\bibliography{jabref}

\begin{thebibliography}{10}
\providecommand{\url}[1]{#1}
\csname url@samestyle\endcsname
\providecommand{\newblock}{\relax}
\providecommand{\bibinfo}[2]{#2}
\providecommand{\BIBentrySTDinterwordspacing}{\spaceskip=0pt\relax}
\providecommand{\BIBentryALTinterwordstretchfactor}{4}
\providecommand{\BIBentryALTinterwordspacing}{\spaceskip=\fontdimen2\font plus
\BIBentryALTinterwordstretchfactor\fontdimen3\font minus
  \fontdimen4\font\relax}
\providecommand{\BIBforeignlanguage}[2]{{%
\expandafter\ifx\csname l@#1\endcsname\relax
\typeout{** WARNING: IEEEtran.bst: No hyphenation pattern has been}%
\typeout{** loaded for the language `#1'. Using the pattern for}%
\typeout{** the default language instead.}%
\else
\language=\csname l@#1\endcsname
\fi
#2}}
\providecommand{\BIBdecl}{\relax}
\BIBdecl

\bibitem{Merwaday2014}
A.~Merwaday, N.~Rupasinghe, Ä.~Güvenç, W.~Saad, and M.~Yuksel, ``{USRP}-based
  indoor channel sounding for {D2D} and multi-hop communications,'' in
  \emph{WAMICON}, 2014, pp. 1--6.

\bibitem{Zelenbaba2020}
S.~Zelenbaba, D.~Löschenbrand, M.~Hofer, A.~Dakić, B.~Rainer, G.~Humer, and
  T.~Zemen, ``A {Scalable} {Mobile} {Multi}-{Node} {Channel} {Sounder},'' in
  \emph{{IEEE} {Wireless} {Communications} and {Networking} {Conference}
  ({WCNC})}, May 2020, pp. 1--6.

\bibitem{Stanko2021}
D.~Stanko, G.~Sommerkorn, A.~Ihlow, and G.~Del~Galdo, ``Enable software-defined
  radios for real-time {MIMO} channel sounding,'' in \emph{IEEE International
  Instrumentation and Measurement Technology Conference (I2MTC)}, 2021, pp.
  1--5.

\bibitem{Zhang2020}
G.~Zhang, X.~Cai, W.~Fan, and G.~F. Pedersen, ``A {USRP}-{Based} {Channel}
  {Sounder} for {UAV} {Communications},'' in \emph{14th {European} {Conference}
  on {Antennas} and {Propagation} ({EuCAP})}, Mar. 2020, pp. 1--4.

\bibitem{8576578}
X.~Cai, J.~Rodríguez-Piñeiro, X.~Yin, N.~Wang, B.~Ai, G.~F. Pedersen, and
  A.~P. Yuste, ``An empirical air-to-ground channel model based on passive
  measurements in {LTE},'' \emph{IEEE Transactions on Vehicular Technology},
  vol.~68, no.~2, pp. 1140--1154, 2019.

\bibitem{9454158}
X.~Cai, T.~Izydorczyk, J.~Rodríguez-Piñeiro, I.~Z. Kovács, J.~Wigard,
  F.~M.~L. Tavares, and P.~E. Mogensen, ``Empirical low-altitude air-to-ground
  spatial channel characterization for cellular networks connectivity,''
  \emph{IEEE Journal on Selected Areas in Communications}, vol.~39, no.~10, pp.
  2975--2991, 2021.

\bibitem{7928479}
X.~Cai, A.~Gonzalez-Plaza, D.~Alonso, L.~Zhang, C.~B. Rodríguez, A.~P. Yuste,
  and X.~Yin, ``Low altitude {UAV} propagation channel modelling,'' in
  \emph{2017 11th European Conference on Antennas and Propagation (EUCAP)},
  2017, pp. 1443--1447.

\bibitem{similar_paper}
B.~Gokalgandhi, P.~Maddala, and I.~Seskar, ``Implementation of fgpa based
  channel sounder for large scale antenna systems using rfnoc on usrp
  platform,'' 09 2017.

\bibitem{rfnoc}
\BIBentryALTinterwordspacing
{Ettus Research}, ``{RFNoC}.'' [Online]. Available:
  \url{https://www.ettus.com/sdr-software/rfnoc/}
\BIBentrySTDinterwordspacing

\bibitem{uhd_github}
\BIBentryALTinterwordspacing
------, ``{USRP Hardware Driver (UHD™) Software}.'' [Online]. Available:
  \url{https://github.com/EttusResearch/uhd}
\BIBentrySTDinterwordspacing

\end{thebibliography}

\end{document}